\documentclass[12pt]{article}
\usepackage{epsf}
\usepackage{cite}
\usepackage{amssymb}
\newcommand{\be}{\begin{equation}}
\newcommand{\ee}{\end{equation}}
\newcommand{\bea}{\begin{eqnarray}}
\newcommand{\eea}{\end{eqnarray}}
\newcommand{\ba}{\begin{array}}
\newcommand{\ea}{\end{array}}

\newcommand{\pl}{Phys.\ Lett.}
\newcommand{\np}{Nucl.\ Phys.}

\newcommand{\prd}{Phys.\ Rev.\ D}

\newcommand{\zp}{Zeit.\ Phys.} 
 
\newcommand{\rmp}{Rev.\ Mod.\ Phys.} 
\newcommand{\etal}{{\it et al}.,\ }

\begin{document}
\title{QCD corrections to $b\rightarrow s\gamma\gamma$ induced decays:\\
$B\rightarrow X_s\gamma\gamma$ and $B_s\rightarrow\gamma\gamma$}
\author{\renewcommand{\thefootnote}{\alph{footnote}}
L. Reina\footnotemark[1], G. Ricciardi\footnotemark[2], 
A. Soni\footnotemark[1]\\
\renewcommand{\thefootnote}{\alph{footnote}}
\footnotemark[1] Physics Department, Brookhaven National Laboratory,\\
 Upton, NY 11973 \\
\renewcommand{\thefootnote}{\alph{footnote}}
\footnotemark[2] Dipartimento di Scienze Fisiche, Universit\`a degli 
Studi di Napoli,\\
and I.N.F.N., Sezione di Napoli,\\
Mostra d'Oltremare, Pad. 19, I-80125 Napoli, Italy }
\date{}
\maketitle
\renewcommand{\thefootnote}{\arabic{footnote}}
\begin{abstract}
  We present a complete calculation of the Leading Order QCD
  corrections to the quark level decay amplitude for $b\rightarrow
  s\gamma\gamma$ and study their relevance for both the inclusive
  branching ratio $BR(B\rightarrow X_s\gamma\gamma)$ and for the
  exclusive decay channel $B_s\rightarrow \gamma\gamma$. In addition
  to the uncertainties in the short distance calculation, due to the
  choice of the renormalization scale, an appreciable uncertainty in
  both $B_s\rightarrow\gamma\gamma$ and $B\rightarrow X_s\gamma\gamma$
  is introduced by the matrix element calculation. We also briefly
  discuss some long distance effects, especially those due to the
  $\eta_c$ resonance for the inclusive rate. Finally, a brief analysis
  of the IR singularities of the two photon spectrum in the inclusive
  case is given.
\end{abstract} 
\baselineskip 20pt

\section{Introduction} 

The radiative decays of the B meson are known to be very sensitive to
strong interaction perturbative corrections as well as to the flavor
structure of the electroweak interactions and to new physics beyond
the Standard Model.  In particular, both inclusive and exclusive
processes induced by $b\rightarrow s\gamma$ have been studied in great
detail \cite{bsgall,misiak,bsglocoeff,savage,voloshin} and two
measurements already exist from the CLEO collaboration:
$BR(B\rightarrow X_s\gamma)\!=\!(2.32\pm 0.57\pm 0.35)\times 10^{-4}$
and $BR(B\rightarrow K^*\gamma)\!=\!(4.2\pm 0.8\pm 0.6)\times
10^{-5}$.

Due to the impressive experimental effort which is being directed to
the study of the physics of the B meson, we can be confident that much
lower branching ratios will be measured in the future.
Therefore it may be interesting to study processes induced at the
quark level by a two photon radiative decay of the b quark, i.e. by
$b\rightarrow s\gamma\gamma$. 

The $b\rightarrow s\gamma\gamma$ decay has received some attention in
the literature \cite{yao,simma,herrlich}, because of the interest in
the $B_s\rightarrow\gamma\gamma$ exclusive mode.  More recently, in
Ref.~\cite{bsgg} we focused on the study of the inclusive
$B\rightarrow X_s\gamma\gamma$ branching ratio.  In the pure
electroweak theory, without QCD corrections but \emph{after the
  necessary kinematical cuts to isolate the contribution into hard
  photons are imposed}, both branching ratios are found to be of order
$10^{-7}$. There is at present an experimental upper bound on the
$BR(B_s\rightarrow\gamma\gamma)$, namely
$BR(B_s\rightarrow\gamma\gamma)<1.48\times 10^{-4}$ \cite{l3}.

As we know from the study of $b\rightarrow s\gamma$, the impact of QCD
corrections on radiative B decays can be pretty dramatic. Therefore in
this sequel, as we promised in \cite{bsgg}, we now want to present the
study of Leading Order QCD corrections to the quark level process
$b\rightarrow s\gamma\gamma$.  We will use this result to predict the
QCD corrected branching ratios for both the inclusive $B\rightarrow
X_s\gamma\gamma$ and the exclusive $B_s\rightarrow\gamma\gamma$ mode.
In both cases QCD corrections increase the branching ratio by $60\%$
to more than $100\%$. On the other hand, the forward-backward
asymmetry that was introduced in \cite{bsgg} turns out to be very
robust with respect to QCD corrections and always varies by less than
$15\%$.

In order to motivate the interest of our perturbative calculation we
will also comment about some relevant long distance contributions and
devote particular attention to the effect of the $\eta_c$ resonance in
the inclusive case. Moreover, we will see how some uncertainty for
both the inclusive and the exclusive branching ratio is introduced at
the level of the matrix element calculation, due to the dependence on
$m_s$.

Finally, we will give in Appendix \ref{ir} the detailed description of
the treatment of the IR singularities which arise in the spectrum of
the two photons for $B\rightarrow X_s\gamma\gamma$.

\section{Leading Order QCD corrections to $\lowercase{b}
\rightarrow\lowercase{s\gamma\gamma}$.}
\label{generalqcd}

In this section we present the general structure of the leading Order
QCD corrections to the quark level decay process $b\rightarrow
s\gamma\gamma$. We will give the expression for the amplitude
$A(b\rightarrow s\gamma\gamma)$, including a complete resummation of
the leading QCD corrections to all orders in
$(\alpha_s\log(\mu^2/M_W^2))^n$. The result will be then specialized
in the following sections to the calculation of the inclusive
branching ratio $BR(B\rightarrow X_s\gamma\gamma)$ and of the
exclusive branching ratio for the decay $B_s\rightarrow
\gamma\gamma$.

We will discuss QCD corrections in the well established framework of
electroweak effective hamiltonians with renormalization group improved
resummation of QCD corrections. For a complete review of the subject
see Ref. \cite{burasreview}.  The most general effective hamiltonian
which describes radiative $b\rightarrow s$ decays with up to three
emitted gluons or photons is given by \cite{grinstein,simmaeom}

\be
\label{heff}
H_{eff} = -\frac{4G_F}{\sqrt{2}} V_{tb} V_{ts}^*\sum_{i=1}^{8}
C_i(\mu) O_i\,\,\,,
\ee

\noindent where, as usual, $G_F$ denotes the Fermi coupling constant
and $V_{ij}$ indicates some Cabibbo-Kobayashi-Maskawa (CKM) matrix
element.  In writing Eq.~(\ref{heff}), we have used the unitarity of the
CKM matrix and we have taken into account that for $b\rightarrow s$
transitions $V_{ub}V_{us}^*\ll V_{tb}V_{ts}^*\simeq V_{cb}V_{cs}^*$.
The basis of local operators we use is obtained from the more general
set of gauge invariant dimension five and six local operators with up
to three external gauge bosons by appling the QED and QCD equations of
motion \cite{grinstein,simmaeom} and is expressed in terms of the
following operators

\bea
\label{operators}
O_1 &=& (\bar s_\alpha\gamma^\mu L c_\beta)(\bar c_\beta\gamma_\mu L 
b_\alpha) \,\,\,,\nonumber\\
O_2 &=& (\bar s_\alpha\gamma^\mu L c_\alpha)(\bar c_\beta\gamma_\mu L 
b_\beta) \,\,\,,\nonumber\\
O_{3,5} &=& (\bar s_\alpha\gamma^\mu L b_\alpha)
  \sum_{q=u,\ldots,b}(\bar q_\beta\gamma_\mu (L,R) q_\beta)\,\,\,, \\
O_{4,6} &=& (\bar s_\alpha\gamma^\mu L b_\beta)
  \sum_{q=u,\ldots,b}(\bar q_\beta\gamma_\mu (L,R) q_\alpha)\,\,\,, 
\nonumber\\
O_7 &=& \frac{e}{16\pi^2}\bar s_\alpha \sigma^{\mu\nu}(m_b R+m_s L)
 b_\alpha F_{\mu\nu} \,\,\,,\nonumber\\
O_8 &=& \frac{g_s}{16\pi^2}\bar s_\alpha \sigma^{\mu\nu}(m_b R+m_s L)
 t^a_{\alpha\beta} b_\beta G^a_{\mu\nu}\,\,\,, \nonumber
\eea

\noindent where the chiral structure is specified by the projectors
$L,R = (1\mp\gamma_5)/2$, while $\alpha$ and $\beta$ are color
indices. $F_{\mu\nu}$ and $G_{\mu\nu}^a$ denotes the QED and QCD
field strength tensors respectively, also $e$ and $g_s$ stand for the
electromagnetic and strong coupling constants.

The Wilson coefficients $C_i(\mu)$ are process independent and their
renormalization is determined only by the basis of operators
$\{O_i\}$.  They depend on the renormalization scale $\mu$ which we
will set eventually to $\mu\approx m_b$. This introduces an error in
the theory that is quite significant when only Leading Order (LO)
logarithms of the form $(\alpha_s\log(\mu^2/M_W^2))^n$ are taken into
account and gets appreciably reduced when also Next-to-Leading Order
(NLO) logarithms of the form $\alpha_s(\alpha_s\log(\mu^2/M_W^2))^n$
are resummed.  The LO result for the Wilson coefficients in
Eq.~(\ref{heff}) is now a well established result \cite{bsgall} and
recently the authors of Ref. \cite{misiak} provided us with the first
NLO calculation.

If we want to calculate the amplitude for $b\rightarrow s\gamma\gamma$
at LO we have to use the effective hamiltonian in Eq.~(\ref{heff})
with LO Wilson coefficients and evaluate its matrix element for the
$b\rightarrow s\gamma\gamma$ decay at $O(\alpha_s^0)$. On the other
hand, for a NLO result we have to use NLO Wilson coefficients and
include $O(\alpha_s)$ corrections to the matrix element.

In order to understand the impact of QCD corrections on this new class
of rare radiative B-decays, we choose to perform our analysis
including, for the time being, only LO corrections. Therefore we will
take the LO regularization scheme independent Wilson coefficients from
the literature \cite{bsglocoeff} and will not consider explicitly the
matrix elements due to the insertion of $O_5$ and $O_6$ into the one
photon and one gluon penguin diagrams. In fact these matrix elements
are reabsorbed into the scheme independent definition of $C_7(\mu)$
and $C_8(\mu)$

\be 
\label{sicoeff}
C_{7,8}^{\rm eff}(\mu)=C_{7,8}(\mu)+\vec Z_{7,8}^T\cdot \vec
C(\mu) \,\,\,,
\ee 

\noindent where $\vec C(\mu)$ is the vector of
$C_1(\mu),\ldots,C_6(\mu)$, while the vectors $\vec Z_{7,8}$ depend on
the regularization scheme: they are zero in the 't Hooft-Veltman (HV)
scheme and non zero in the Na\"{\i}ve Dimensional Reduction scheme
(NDR) (see Ref.~\cite{bsglocoeff} for details). In our calculation, we
use the $C_i^{\rm eff}$ effective coefficients, although we decide to
drop the extra index to simplify the notation. We note that no new
regularization scheme dependence enters in the calculation of the
matrix elements for $b\rightarrow s\gamma\gamma$ through the new class
of penguin diagrams with two external photons.  In fact, a finite
scheme dependence in the matrix element can arise only as a result of
the product of the UV pole part of a Feynman diagram (or set of
diagrams) times some $O(\epsilon)$ evanescent Dirac structure of the
diagram itself.  However, as we will see, the new penguins with two
external photons are UV finite at $O(\alpha_s^0)$. Therefore any
difference between two regularization schemes can only give an
unphysical $O(\epsilon)$ effect.  We have performed the calculation of
the following matrix elements in both the HV and NDR regularization
schemes and, as expected, the results coincide. Therefore we do not
specify any regularization scheme in the following discussion.

\begin{figure}[t]
\centering
\epsfxsize=6.in
\leavevmode\epsffile{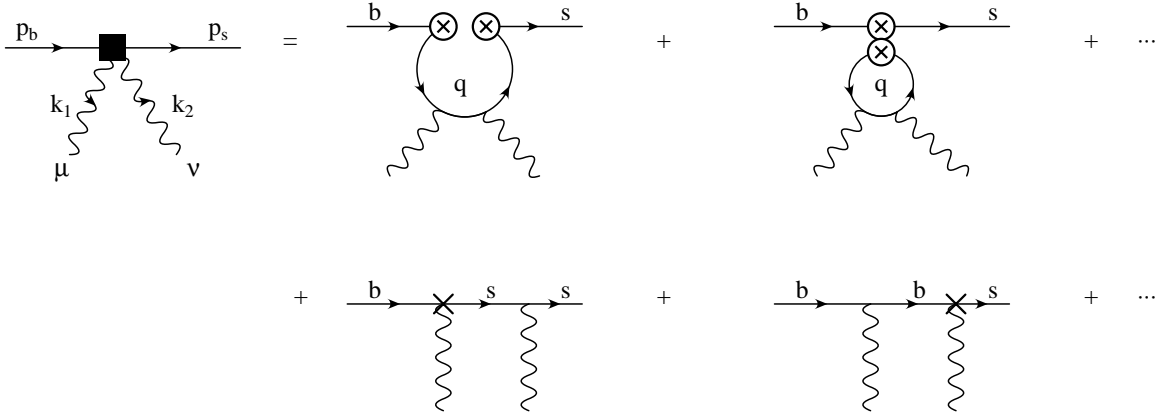}
\caption[]{Examples of Feynman diagrams which contribute to the matrix
element $\langle s\gamma\gamma|H_\mathrm{eff}|b\rangle$. The 1PI
diagrams illustrate the two possible insertions of the operators
$O_1,\dots,O_6$ (double circled cross vertices), depending on their
flavor, chiral and color structure, while the 1PR ones represent the
insertion of $O_7$ (cross vertices). Moreover \emph{q} indicates a
generic quark flavor.}
\label{bsggeff}
\end{figure}

The amplitude for the decay $b(p)\rightarrow
s(p^\prime)+\gamma(k_1)+\gamma(k_2)$ can be expressed as

\be
\label{ampl}
A = \sum_{i=1}^{7} A_i = -\frac{i e^2G_F}{\sqrt{2}\pi^2}\lambda_t
  \sum_{i=1}^{7}C_i(\mu)\bar u_s(p^\prime) T_i^{\mu\nu}u_b(p)
  \epsilon_\mu(k_1)\epsilon_\nu(k_2)\,\,\,,
\ee

\noindent where $\lambda_t=V_{tb}V_{ts}^*$ and $\epsilon_\mu(k_1)$ and
$\epsilon_\nu(k_2)$ are the polarization vectors of the two
photons. The $C_i(\mu)$ coefficient are intended to be the LO ones, as
explained before, while we have denoted by $T_i^{\mu\nu}$ the tensor
structure of the transition amplitude induced by the operator $O_i$.
The different $T_i^{\mu\nu}$ are obtained inserting the operators of
Eq.~(\ref{operators}) into the Feynman diagrams of Fig.~\ref{bsggeff},
according to the color and chiral structure of the operators
themselves.
\noindent In particular, one has to be careful when dealing with 
penguin-like operators $O_{3,\ldots,6}$ due to their more complicated
flavor structure. The $T_i^{\mu\nu}$ tensors can be summarized in a
compact form as follows

\bea
\label{tmunu}
T_1^{\mu\nu} &=& N_c Q_u^2 \kappa_c W_2^{\mu\nu}\,\,\,,\nonumber\\
T_2^{\mu\nu} &=& Q_u^2 \kappa_c W_2^{\mu\nu}\,\,\,,\nonumber\\
T_3^{\mu\nu} &=& \left\{N_c\left[
    Q_d^2\left(\kappa_d+\kappa_s+\kappa_b\right)+
    Q_u^2\left(\kappa_u+\kappa_c\right)\right]+
  Q_d^2\left(\kappa_b+\kappa_s\right)\right\} W_2^{\mu\nu}\,\,\,,
\nonumber\\
T_4^{\mu\nu} &=& \left\{\left[
    Q_d^2\left(\kappa_d+\kappa_s+\kappa_b\right)+
    Q_u^2\left(\kappa_u+\kappa_c\right)\right]+
  N_c Q_d^2\left(\kappa_s+\kappa_b\right)\right\}W_2^{\mu\nu}\,\,\,,\\
T_5^{\mu\nu} &=& -N_c\left[
  Q_d^2\left(\kappa_d+\kappa_s+\kappa_b\right)+
  Q_u^2\left(\kappa_u+\kappa_c\right)\right] W_2^{\mu\nu}+
Q_d^2\left[m_b W_{5,b}^{\mu\nu} R+m_s W_{5,s}^{\mu\nu} L\right]
\,\,\,,\nonumber\\
T_6^{\mu\nu} &=& -\left[ Q_d^2\left(\kappa_d+\kappa_s+\kappa_b\right)+
  Q_u^2\left(\kappa_u+\kappa_c\right)\right] W_2^{\mu\nu}+
N_c Q_d^2\left[m_b W_{5,b}^{\mu\nu} R+m_s W_{5,s}^{\mu\nu} L\right]
\,\,\,,\nonumber\\
T_7^{\mu\nu} &=& Q_d W_7^{\mu\nu}\,\,\,,\nonumber \eea

\noindent where we note that there is no contribution from the
chromo-magnetic operator $O_8$ at $O(\alpha_s^0)$.  In
Eq.~(\ref{tmunu}) $N_c$ denotes the number of colors ($N_c\!=\!3$),
$Q_u\!=\!2/3$ and $Q_d\!=\!-1/3$ are the up-type and down-type quark
electric charges and $m_b$ and $m_s$ indicate the masses of the bottom
and of the strange quark respectively. Moreover all the $T_i^{\mu\nu}$
have been expressed in terms of only three tensor structures

\bea
\label{wtensors}
W_2^{\mu\nu} &=& \left\{\frac{1}{k_1\cdot k_2}\left[
k_1^\nu k_1\!\!\!\!/\gamma^\mu k_2\!\!\!\!/-
k_2^\mu k_1\!\!\!\!/\gamma^\nu k_2\!\!\!\!/- 
k_2^\mu k_1^\nu \left(k_1\!\!\!\!/-k_2\!\!\!\!/\right)\right]+\right.
 \nonumber\\
&& \,\,\,\,\,\,\,\,\,
\left.\phantom{\frac{1}{k_1\cdot k_2}}
\gamma^\nu\gamma^\mu\left(k_1\!\!\!\!/-k_2\!\!\!\!/\right)-
g^{\mu\nu}\left(k_1\!\!\!\!/-k_2\!\!\!\!/\right)+
2k_1^\nu\gamma^\mu\right\} L \,\,\,,\\
W_{5,q}^{\mu\nu} &=& \frac{1}{m_q^2}\left(
\gamma^\nu k_2\!\!\!\!/\gamma^\mu k_1\!\!\!\!/ + 
k_1\cdot k_2\gamma^\nu\gamma^\mu+g^{\mu\nu}k_2\!\!\!\!/\, k_1\!\!\!\!/-
k_2^\mu\gamma^\nu k_1\!\!\!\!/-k_1^\nu k_2\!\!\!\!/\gamma^\mu\right)
\left(1-2\kappa_q\right)+\nonumber\\
&& \,\,\,\,\,\,\,\,\, 4\left(g^{\mu\nu}-\frac{k_1^\nu k_2^\mu}
{k_1\cdot k_2}\right)\kappa_q \,\,\,, \nonumber\\
W_7^{\mu\nu} &=& \frac{1}{2}\left[
-\frac{1}{2p\cdot k_2}k_1\!\!\!\!/\gamma^\mu(m_b R+m_s L)
(p\!\!\!/-k_2\!\!\!\!/+m_b)\gamma^\nu+ \right. \nonumber\\
&&\,\,\,\,\,\,\,\left.
\frac{1}{2p^\prime\cdot k_2}\gamma^\nu(p\!\!\!/-k_1\!\!\!\!/+m_s)
k_1\!\!\!\!/\gamma^\mu(m_b R+m_s L)\right]\,+\,\left(\{k_1,\mu\}
\leftrightarrow\{k_2,\nu\}\right)\,\,\,, \nonumber 
\eea

\noindent and the analytic coefficients $\kappa_q$ defined as

\be
\kappa_q=\frac{1}{2}+\frac{Q_0(z_q)}{z_q}=
\frac{1}{2}+\frac{1}{z_q}\int_0^1\frac{dx}{x}\log(1-z_q x+z_qx^2)\,\,\,,
\label{kappa}
\ee

\noindent for $z_q\!=\!2k_1\cdot k_2/m_q^2\,\,$. In derivng
Eq.~(\ref{tmunu})-(\ref{kappa}) we have checked the analogous results
given in Refs.~\cite{aligreub,pott} for the $b\rightarrow s\gamma g$
decay and we confirm all of them.

Finally, we observe that using the effective hamiltonian of
Eq.~(\ref{heff}) at $\mu\simeq M_W$ and in the absence of QCD
corrections, we can reproduce the pure electroweak amplitude obtained
in Refs.~\cite{yao,simma,herrlich}, as expected. Only two operators,
$O_2$ and $O_7$, contribute in this case. Their Wilson coeffcients at
$\mu\!=\!M_W$ read

\bea
\label{pureewic}
C_2(M_W) &=& 1\,\,\,,\nonumber \\
C_7(M_W) &=& F_2(x_t)-F_2(x_c) \,\,\,,
\eea

\noindent where $F_2(x_i)$ is the Inami-Lim function for the
on-shell $bs\gamma$ vertex \cite{inamilim}

\be
\label{f2}
F_2(x_i) = \frac{3x_i^3-2x_i^2}{4(x_i-1)^4}\log x_i+
           \frac{-8x_i^3-5x_i^2+7x_i}{24(x_i-1)^3}\,\,\,.
\ee

\noindent The corresponding matrix elements are given in
Eq.~(\ref{tmunu}) and we can easily verify that $O_2$ reproduces the
one particle irreducible part of the result of
Refs.~\cite{yao,simma,herrlich} while $O_7$ is responsible for the one
particle reducible part.

\section{Inclusive branching ratio for $\lowercase{b}\rightarrow 
\lowercase{s\gamma\gamma}$.}
\label{inclusive}

As already discussed in Ref.~\cite{bsgg}, the inclusive rate for
$B\rightarrow X_s\gamma\gamma$ can be described to a good degree of
accuracy by the quark level process. We can therefore directly use the
results of Section \ref{generalqcd} to evaluate the square amplitude.
For this purpose, we rewrite the amplitude as

\be
\label{amplincl}
    A = -\frac{i e^2G_F}{\sqrt{2}\pi^2}\lambda_t \bar u_s(p^\prime)
\left[ F_2 W_2^{\mu\nu} + 
       F_5 \left(m_b W_{5,b}^{\mu\nu}R+m_s W_{5,s}^{\mu\nu}L\right)+ 
       F_7 W_7^{\mu\nu}\right]
       u_b(p)\epsilon_\mu(k_1)\epsilon_\nu(k_2)\,\,\,,
\ee

\noindent where the coefficients $F_i$ can be easily deduced from
Eqs.~(\ref{ampl})-(\ref{tmunu}), and are 

\bea 
\label{fcoeff} 
  F_2 &=& (N_c C_1(\mu)+C_2(\mu))Q_u^2\kappa_c+\nonumber\\
   &&  C_3(\mu)\left\{
       N_c\left[Q_d^2(\kappa_d+\kappa_s+\kappa_b)+
        Q_u^2\left(\kappa_u+\kappa_c\right)\right]+
        Q_d^2(\kappa_s+\kappa_b)\right\}+\nonumber\\ 
   &&  C_4(\mu)\left\{\left[Q_d^2(\kappa_d+\kappa_s+\kappa_b)+
       Q_u^2\left(\kappa_u+\kappa_c\right)\right]+
       N_c Q_d^2\left(\kappa_s+\kappa_b\right)\right\}-\nonumber\\ 
   &&  (N_c C_5(\mu)+C_6(\mu))\left[
       Q_d^2(\kappa_d+\kappa_s+\kappa_b)+
       Q_u^2\left(\kappa_u+\kappa_c\right)\right]\,\,\,,\\ 
   F_5 &=& (C_5(\mu)+N_c C_6(\mu))Q_d^2\,\,\,,\nonumber\\ 
   F_7 &=& C_7(\mu) Q_d\,\,\,.\nonumber 
\eea

\noindent The square amplitude summed over spins and polarizations
will then be given by

\bea
\label{msqr} 
   |A|^2 &=& \frac{1}{4} \left(\frac{e^2G_F}{\sqrt{2}\pi^2} 
   \lambda_t\right)^2 m_b^4 \left[
   |F_2|^2 A_{22}+|F_5|^2 A_{55}+|F_7|^2 A_{77}+ 
      2 \mbox{Re}(F_7 F_2^*) A_{27}+\right.\\ 
    && \left.  2\mbox{Re}(F_5 F_2^*(1-2\kappa_b)) A_{25}^b+ 
               2\mbox{Re}(F_5 F_2^* (1-2\kappa_s)) A_{25}^s+ 
          2\mbox{Re}(F_7 F_5^*) A_{57}\right]\nonumber \,\,\,,
\eea

\noindent where the quantities $A_{ij}$ denote the contractions between
the tensors $W_i^{\mu\nu}$ and $W_j^{\mu\nu}$. In order to give them
explicitily we introduce the following notation\footnote{We decide to
follow in our discussion the notation of Ref.~\cite{pott} as closely
as possible, which can be helpful for comparison.}

\be
\label{invariants}
 s = \frac{2k_1\cdot k_2}{m_b^2}\,\,\,\,,\,\,\,\,
 t = \frac{2p\cdot k_2}{m_b^2}\,\,\,\,,\,\,\,\,
 u = \frac{2p\cdot k_1}{m_b^2}\,\,\,\,,\,\,\,\,
 \rho = \frac{m_s^2}{m_b^2}\,\,\,,
\ee

\noindent which satisfy the relation: $u+t-s=1-\rho$. In order to 
introduce a more compact notation, it can be useful to switch
occasionaly to the ($\bar s$, $\bar u$, $\bar t$) invariants, defined
as $\bar s\!=\!s/(1-\rho)$, $\bar t\!=\!t/(1-\rho)$ and $\bar
u\!=\!u/(1-\rho)$. In this framework the $A_{ij}$ quantities are given
by:

\bea
\label{aij}
A_{22} &=& 2\left[(1-\rho)^2-(1+\rho)s\right]\,\,\,,\nonumber\\
A_{55} &=& \left[16|\kappa_b|^2+|(1-2\kappa_b)s+4\kappa_b|^2+
 \rho\left(16|\kappa_s|^2+|(1-2\kappa_s)s/\rho+4\kappa_s|^2\right)
 \right](1-s+\rho)+\nonumber\\
 && 16\,\mbox{Re}\left\{8\rho\kappa_b\kappa_s^*+s\left[\kappa_b-
     2(1+\rho)\kappa_b\kappa_s^*+\rho\kappa_s^*\right]\right\}\,\,\,,
     \nonumber\\
A_{25}^{b,s} &=& \pm s(1-\rho\mp s)\,\,\,,\\
A_{27} &=& -2\left[(1+\rho)s+\frac{\rho s^2}{(s-t)t}+
            \frac{\rho s^2}{(s-u)u}\right]\,\,\,,\nonumber\\
A_{57} &=& \mbox{Re}\left\{\phantom{\frac{1}{2}}8(\kappa_b+
\rho\kappa_s)s -
\left[4\rho(\kappa_b+\kappa_s)+s\left((1-2\kappa_s)+\rho(1-2\kappa_b)
    \right)\right]\right.\nonumber\\
    && \left.\left[\frac{s^2}{t(s-t)}+\frac{s^2}{u(s-u)}\right]\right\}
       \,\,\,,\nonumber\\
A_{77} &=& (1+\rho)\left[(1-\rho)A_{77}^{(1)}-2A_{77}^{(2)}\right]+
       A_{77}^{(3)}\,\,\,,\nonumber
\eea

\noindent with

\bea
\label{a77}
A_{77}^{(1)} &=& \frac{1}{\bar t}\left[1+\bar u+
   \frac{2\bar u(\bar u-2)}{1-\bar u}\bar t+
   \frac{2\bar u-1}{1-\bar u}\bar t^2\right] + 
   (\bar t\leftrightarrow \bar u)\,\,\,,\nonumber\\
A_{77}^{(2)} &=& \frac{1}{\bar t^2}\left[1-
   \frac{1+\rho}{1-\bar u}\bar t+
   \frac{\rho}{(1-\bar u)^2}\bar t^2\right] +
   (\bar t\leftrightarrow \bar u)\,\,\,,\\
A_{77}^{(3)} &=& -2\frac{s}{\bar t \bar u}\left\{(1+\rho)(2+\bar u 
\bar t)+
   \frac{\rho}{1-\rho}\left[1-\frac{2(1+\rho)-\bar t \bar u}
  {(1-\bar t)(1-\bar u)}\right]\bar s\right\}\,\,\,.\nonumber
\eea

\noindent We want to put particular emphasis on the structure of the
$A_{77}$ part of the square amplitude because it will be a crucial
ingredient in testing the cancellation of the IR divergences
which appear in the calculation of the total rate.
In fact, the total rate is obtained by integrating 

\be 
\label{dgamma}
d\Gamma = \frac{1}{2 m_b (2\pi)^{2D-3}}\delta^D(p-p^\prime-k_1-k_2)|A|^2
  \frac{d^{(D-1)}p^\prime}{2p^\prime_0}\frac{d^{(D-1)}k_1}{2\omega_1}
  \frac{d^{(D-1)}k_2}{2\omega_2}\,\,\,,
\ee

\noindent over the physical phase space, where we have denoted by
$\omega_1$ and $\omega_2$ the energies of the two photons. All the
terms in $|A|^2$ are both UV and IR finite except $A_{77}$ which gives
origin to IR singularities upon integration over the phase space of
the two photons. We chose to regularize the integrals working in
$D=4-2\epsilon$ dimensions and to extract the existing IR
singularities as poles in $1/\epsilon$. These IR divergences originate
when either $\omega_1\rightarrow 0$ or $\omega_2\rightarrow 0$, and
correspond to the well known IR singularities which arise in the
bremstrahlung process when one or the other of the two photons becomes
very soft\footnote{We note that there are no collinear singularities
so long as the mass of the external quarks are non zero. This gives
origin to a non negligible dependence on $m_s$ and perhaps a more
careful resummation of logarithms like $\log(m_s^2)$ in the rate
should be implemented. We will discuss our concern with this problem
later on.}. In this limit the $b\rightarrow s\gamma\gamma$ decay
cannot be distinguished from $b\rightarrow s\gamma$ and the two
processes have to be considered together in order to obtain meaningful
(i.e., finite) physical quantities. In fact, we have checked that the
IR singularities which arise from the integration of $A_{77}$ over the
phase space cancel exactly with the $O(\alpha_e)$ virtual corrections
to the $b\rightarrow s\gamma$ amplitude (see
Appendix~\ref{ir}). Therefore $(\Gamma(b\rightarrow s\gamma)+
\Gamma(b\rightarrow s\gamma\gamma))$ is free of IR singularities.

This problem has already been studied in detail in order to take into
account the $O(\alpha_s)$ bremstrahlung corrections for $b\rightarrow
s\gamma$ \cite{aligreub,pott}. However our point of view here is
slightly different. In our case the bremstrahlung process is not
considered as an $O(\alpha_e)$ correction to the $b\rightarrow
s\gamma$ amplitude, but as a different process: the decay of a $b$
quark into an $s$ quark plus two hard photons. Therefore, the
endpoints of the spectrum of each photon (where the IR singularities
are present) do not in fact correspond to the process of interest. In
order to calculate the physical rate of interest we just have to
impose a cut on the energy of each photon, which will naturally
correspond to the experimental cut imposed on the minimun energy for
detectable photons.

\begin{figure}[ht]
\centering
\epsfxsize=4.in
\leavevmode\epsffile{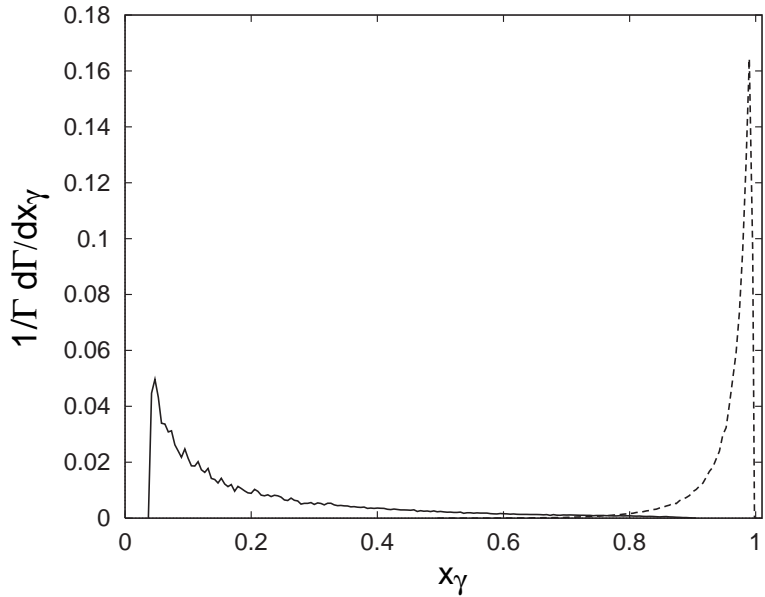}
\caption[]{The spectrum of the two photons including QCD corrections,
normalized to the total QCD corrected rate for $m_s\!=\!0.5$ GeV.
The two photons are defined as the photon of lower energy (solid) and
the photon of higher energy (dashed).}
\label{spectrum}
\end{figure}

In Fig.~\ref{spectrum} we illustrate the spectrum of the two photons,
defined as the photon of higher energy and the photon of lower energy.
We obtain this spectrum requiring the energy of each photon to be
larger than $E_\gamma^{\rm min}\!=\!100$ MeV and the angles between
any two outgoing particles to be bigger than $\theta_{ij}^{\rm
min}\!=\!20^\circ$.  This last constraint is not required
analytically, but we think it is reasonable to exclude photons which
are emitted too close to each other or to the outgoing \emph{s} quark,
in order to roughly incorporate the experimental requirements as we
perceive them. Once the structure of the differential rate has been
checked and the presence of IR singularities understood and treated,
we can integrate Eq.~(\ref{dgamma}) numerically and study the impact
of QCD corrections on the total rate as well as on different
distributions.

\noindent We find that QCD corrections enhance the rate by a factor of
$\simeq 2-2.5$, depending on the numerical parameters we use. In our
evaluation we fix $m_b\!=\!4.8$ GeV, $m_c\!=\!1.5$ GeV, $m_t\!=\!
175$ GeV and $|\lambda_t|\!=\!|V_{tb}V_{ts}^*|\!=\!0.04$.  As far as
$m_s$ is concerned, we use $m_s\!\simeq\! M_K\!=\!0.5$ GeV. For this
set of parameters and fixing $\mu\!=\!m_b$, the branching ratio for
$B\rightarrow X_s\gamma\gamma$ goes from $\sim 1.7\times 10^{-7}$,
without QCD corrections, to $\sim 3.7\times 10^{-7}$ when LO QCD
corrections are included. We recall that we define the
$BR(B\rightarrow X_s\gamma\gamma)$ in terms of the semileptonic
branching ratio as follows

\be
\label{brincl}
BR(B\rightarrow X_s\gamma\gamma)\simeq \left[\frac{\Gamma(b\rightarrow
s\gamma\gamma)} {\Gamma(b\rightarrow
cl\nu_l)}\right]^\mathrm{th}\times BR(B\rightarrow X_c
l\nu_l)^\mathrm{exp}\,\,\,,
\ee
\noindent where no QCD corrections have to be included in the
theoretical prediction of $\Gamma(b\rightarrow cl\nu_l)$ at this order
in $\alpha_s$ and we have used $BR(B\rightarrow X_c l\nu_l)\simeq
0.11$ \cite{pdb}. 

In principle, $m_s\!\simeq M_K$ should be used in the phase space
integration, while in the perturbative calculation of the amplitude
one may need to replace it by the current mass $m_s\!\simeq\!0.15$
GeV. However this introduces spurious instabilities in the numerical
Montecarlo integration over the phase space. Since the numerical
results change little as we replace $m_s$ over the range
$0.15\!-\!0.5$ GeV, we prefer to use the same value of $m_s$ in both
cases. Thus, in order to simulate the physical phase space correctly
we set $m_s\!\simeq\!M_K$ everywhere.

Moreover, we have to account for the scale dependence introduced by
QCD corrections at the level of the Wilson coefficients. This makes a
$25-30\%$ uncertainty, as is the case for $B\rightarrow X_s\gamma$
case. For the sake of completion, we also give the values of the
Wilson coefficients we use in Table~\ref{tablecoeff}, for three values
of $\mu$, $\mu\!=\!m_b/2$, $m_b$ and $2m_b$ respectively, and for
$m_t\!=\!175$ GeV and $\alpha_s(M_W)\!=\!0.118$. We will comment about
further uncertainties introduced by long distance QCD effects in
Section \ref{longdistance}.

\begin{table}[ht]
\begin{center}
\begin{tabular}{|l||c c c c c c c|}
\hline\hline
$\mu$ & $C_1$ & $C_2$ & $C_3$ & $C_4$ & $C_5$ & $C_6$ & $C_7$ \\\hline
$m_b/2$ & -0.324 & 1.148 & 0.015 & -0.033 & 0.009 & -0.043 & -0.344 
\\ \hline 
$m_b$ & -0.234 & 1.100 & 0.010 & -0.024 & 0.007 & -0.029 & -0.308 
\\ \hline 
$2m_b$ & -0.162 & 1.065 & 0.007 & -0.017 & 0.005 & -0.019 & -0.277 \\
\hline\hline
\end{tabular}
\caption[]{Values of the regularization scheme independent LO Wilson
coefficients for $\mu\!=m_b/2$, $\mu\!=m_b$, and $\mu\!=2 m_b$, for
$m_b\!=\!4.8$ GeV, $m_t\!=\! 175$ GeV and $\alpha_s(M_W)\!=\!0.118$.}
\label{tablecoeff}
\end{center}
\end{table}

In order to better understand the dynamics of QCD corrections, let us
classify the different contributions to the rate into one particle
reducible (1PR) and one particle irreducible (1PI), as we did in
Ref.~\cite{bsgg} for the pure electroweak case. In the language of the
effective hamiltonian of Eq.~(\ref{heff}) this corresponds to
separating the cotribution of $O_7$ (which corresponds to the IPR
part) from that of all the other operators. As we saw \cite{bsgg}, the
photon invariant mass distribution, $d\Gamma/ds$ is dominated for low
$s$ by the 1PR diagrams, while for larger $s$ a non trivial
$m_c$-dependent contribution from the 1PI diagrams starts being
relevant.  The effect of QCD corrections is to enhance even more the
effect of $O_7$, as we could expect from the dramatic effect that QCD
corrections have in $b\rightarrow s\gamma$, while lowering the impact
of the 1PI contribution, because of the new mixing with many different
four-quark operators. In particular, the contribution of $O_2$ is
suppressed by the destructive interference with $O_1$. We verified
that the contributions of different operators to the angular
distribution of the two photons are very similar to each other, also
after QCD corrections have been included.

On the other hand, as expected, the forward-backward asymmetry we
introduced in \cite{bsgg}

\be
A_\mathrm{FB}=\frac{\Gamma(\cos\theta_{s\gamma}\ge
0)-\Gamma(\cos\theta_{s\gamma}<0)} {\Gamma(\cos\theta_{s\gamma}\ge
0)+\Gamma(\cos\theta_{s\gamma}<0)}\,\,\,\,,
\label{asym}
\ee

\noindent where $\theta_{s\gamma}$ is the angle between the \emph{s}
quark and the softer photon, is rather insensitive to QCD corrections,
since the QCD corrections tend to cancel between the numerator and the
denominator. In fact we find that QCD corrections affect
$A_\mathrm{FB}$ by no more than $15\%$, changing it from 0.71 (without
QCD corrections) to 0.78 (with LO QCD corrections), despite the fact
that the total rate changes by as much as $60\%$ to
$100\%$. Furthermore $A_\mathrm{FB}$ is practically insensitive to the
choice of scale in the LO Wilson coefficients, while the branching
ratio varies as much as $30\%$ with $\mu$.  On the other hand, this
observable will clearly benefit from the enhancement induced by QCD at
the rate level.  Once the process is measured the possibility of
measuring this new observable should give us another handle in testing
our understanding of the theory and in differentiating the Standard
Model from its extensions as already explained in \cite{bsgg}.

\section{The exclusive decay $B_s\rightarrow\gamma\gamma$.}
\label{exclusive}

Using the quark level amplitude in Eq.(\ref{amplincl}) we can also
estimate the rate for the $B_s\rightarrow\gamma\gamma$ rare decay and
evaluate the impact of QCD corrections on it. In order to calculate
the matrix element of (\ref{amplincl}) for the
$B_s\rightarrow\gamma\gamma$ decay, we can work, for instance, in the
weak binding approximation and assume that both the \emph{b} and the
\emph{s} quarks are at rest in the $B_s$ meson. In the rest frame of
the decaying $B_s$ meson we would have that

\be
\label{staticquarks}
k_1\cdot k_2=\frac{M_{B_s}^2}{2}\,\,\,\,\,,\,\,\,\,\, 
p_b\cdot k_1=p_b\cdot k_2 =\frac{1}{2}m_bM_{B_s}\,\,\,\,\,,\,\,\,\,\, 
p_s\cdot k_1=p_s\cdot k_2 =\frac{1}{2}m_sM_{B_s}\,\,\,\,\,.
\ee

\noindent where $m_b$ and $m_s$ must now be traeted as constituent
masses. The problem can also be rephrased in the language of Heavy
Quark Effective Theory (HQET), assuming that the velocity of the
\emph{b} quark coincides with the velocity of the $B_s$ meson up to a
residual momentum of order $\Lambda_\mathrm{QCD}$, i.e.
$p_b^\mu=m_bv^\mu+k^\mu$. To first approximation, the scalar products
of Eq.~(\ref{staticquarks}) are replaced by

\be
\label{hqet}
k_1\cdot k_2=\frac{M_{B_s}^2}{2}\,\,\,\,\,,\,\,\,\,\, 
p_b\cdot k_1=p_b\cdot k_2 =\frac{1}{2}m_bM_{B_s}\,\,\,\,\,,\,\,\,\,\, 
p_s\cdot k_1=p_s\cdot k_2 =\frac{1}{2}(M_{B_s}-m_b)M_{B_s}\,\,\,\,\,,
\ee

\noindent where we have used that $p_s^\mu\!=\!-(p_b-k_1-k_2)^\mu$.
We can see that, to this order, Eqs.~(\ref{staticquarks}) and
(\ref{hqet}) are compatible up to corrections of order
$(\Lambda_\mathrm{QCD}/m_b)$, if we assume $m_s\approx
(M_{B_s}-m_b)\approx\Lambda_\mathrm{QCD}$.  Unless the HQET formalism
is taken to beyond the leading order one cannot make a reliable
distinction between the two predictions. Therefore, for concreteness,
we give in the following the necessary matrix elements using the weak
binding approximation.  By further recalling that

\bea
\langle 0|\bar s\gamma^\mu\gamma_5 b|B_s(P_{B_s})\rangle &=& -i f_{B_s}
P_{B_s}^\mu\,\,\,,\\
\langle 0|\bar s\gamma_5 b|B_s\rangle &=& i f_{B_s}M_{B_s}\,\,\,,
\nonumber
\eea

\noindent we obtain the following matrix elements for $W_2^{\mu\nu}$,
$W_{5,q}^{\mu\nu}$ and $W_7^{\mu\nu}$ 

\bea
\label{bsggmatrixelements}
\langle 0|W_2^{\mu\nu}|B_s\rangle\epsilon_\mu(k_1)\epsilon_\nu(k_2) 
&=& i\frac{1}{2} f_{B_s}
\left(-iF_{\mu\nu} \tilde F^{\mu\nu}\right)\,\,\,\\
\langle 0|W_{5,b}^{\mu\nu}|B_s\rangle\epsilon_\mu(k_1)\epsilon_\nu(k_2) 
 &=& i\frac{1}{4}f_{B_s}M_{B_s}
\left[-\left(\frac{1-2\kappa_b}{m_b}+\frac{8\kappa_b m_b}{M_{B_s}^2}
 \right)F_{\mu\nu}F^{\mu\nu}-
\frac{1-2\kappa_b}{m_b}iF_{\mu\nu}\tilde F^{\mu\nu}\right]\,\,\,,
\nonumber\\
\langle 0|W_{5,s}^{\mu\nu}|B_s\rangle\epsilon_\mu(k_1)\epsilon_\nu(k_2) 
 &=& i\frac{1}{4}f_{B_s}M_{B_s}
\left[\left(\frac{1-2\kappa_s}{m_s}+\frac{8\kappa_s m_s}
 {M_{B_s}^2}\right) F_{\mu\nu}F^{\mu\nu}-
\frac{1-2\kappa_s}{m_s}iF_{\mu\nu}\tilde F^{\mu\nu}\right]\,\,\,,
\nonumber\\
\langle 0|W_7^{\mu\nu}|B_s\rangle\epsilon_\mu(k_1)\epsilon_\nu(k_2) 
&=& i\frac{1}{4}f_{B_s}\frac{(m_b+m_s)^2}{m_bm_s}\left[
\frac{(m_b-m_s)}{(m_b+m_s)}F_{\mu\nu}F^{\mu\nu}-
 i F_{\mu\nu}\tilde F^{\mu\nu}\right]\,\,\,,\nonumber
\eea

\noindent where $f_{B_s}$ denotes the $B_s$ meson decay constant.  The
amplitude $A(B_s\rightarrow\gamma\gamma)$ can therefore be expressed
in terms of the only two tensor structures $F_{\mu\nu}F^{\mu\nu}$ and
$F_{\mu\nu}\tilde F^{\mu\nu}$

\be
\label{b_samplitude}
A(B_s\rightarrow\gamma\gamma)=-i\frac{G_Fe^2}{\sqrt{2}\pi^2}\lambda_t
\left(A^+F_{\mu\nu}F^{\mu\nu}+i\,A^-F_{\mu\nu}\tilde F^{\mu\nu}\right)
\,\,\,,
\ee

\noindent where $\tilde
F^{\mu\nu}\!=\!1/2\,\epsilon^{\mu\nu\rho\sigma} F_{\rho\sigma}$.  The
coefficient $A^+$ and $A^-$ of the CP-even and of the CP-odd term can
be easely derived from Eq.(\ref{bsggmatrixelements}) and read

\bea
\label{apam}
A^+ &=& i\frac{1}{4}f_{B_s}\left[
 -M_{B_s}\left(\frac{1-2\kappa_b}{m_b}+\frac{8\kappa_b m_b}{M_{B_s}^2}
    \right) F_5 
 +M_{B_s}\left(\frac{1-2\kappa_s}{m_s}+\frac{8\kappa_s m_s}{M_{B_s}^2}
    \right) F_5  \right.\nonumber\\
&&  \left.\;\;\;\;\;\;\;\;\;\;\;\;+ \frac{(m_b^2-m_s^2)}{m_b m_s} F_7
    \right]\,\,\,\\
A^- &=& i\frac{1}{4}f_{B_s}\left[
 -2F_2
 -M_{B_s}\left(\frac{1-2\kappa_b}{m_b}+
               \frac{1-2\kappa_s}{m_s}\right)F_5
 -\frac{(m_b+m_s)^2}{m_b m_s}F_7\right]\,\,\,.
\eea

\noindent The QCD corrected coefficients $F_2$, $F_5$ and $F_7$ can be
taken from Eq.~(\ref{fcoeff}), while at $O(\alpha_s^0)$ they are
simply given by $F_2=Q_u^2\kappa_c C_2(M_W)$, $F_5\!=\!0$ and
$F_7\!=\!Q_d C_7(M_W)$, for $C_2(M_W)$ and $C_7(M_W)$ in
(\ref{pureewic}). We notice that the terms proportional to $F_7$ in
both $A^+$ and $A^-$ are inversely proportional to $m_s$\footnote{The
  matrix element of $W_{5,s}^{\mu\nu}$ does not scale as $1/m_s$ for
  small $m_s$ because also $\kappa_s\rightarrow 1/2$ for
  $m_s\rightarrow 0$, therefore killing the $1/m_s$ terms. Moreover,
  the dependence on $m_s$ from this matrix element is very much
  suppressed by the smallness of the coefficient $F_5$.}. This is a
clear signal of the relevance of non-perturbative effects to the
evaluation of the matrix element for the decay rate of
$B_s\rightarrow\gamma\gamma$. In the absence of a calculation of the
matrix elements for this process which takes into account the higher
order corrections in the HQET expansion, we can only give the
perturbative prediction and try to estimate the theoretical error we
have on that. Therefore we will use Eqs.~(\ref{b_samplitude}) and
(\ref{apam}) and vary $m_s$ in the range $300\le m_s\le 500$ MeV.

Let us first estimate the impact of QCD corrections on the rate

\be
\Gamma(B_s\rightarrow\gamma\gamma)=\frac{M_{B_s}^3}{16\pi}
(-i\frac{G_Fe^2}{\sqrt{2}\pi^2}\lambda_t)^2
\left(|A^+|^2+|A^-|^2\right)\,\,\,,
\ee

\noindent and on the ratio of the two coefficients $A^+$ and $A^-$

\be
R = \frac{|A^+|^2}{|A^-|^2}\,\,\,.
\ee

\noindent As pointed out in Refs.~\cite{yao,herrlich}, 
the coefficients $A^+$ and $A^-$ correspond respectively to photons
with parallel ($\epsilon(k_1)\cdot\epsilon(k_2)$) and perpendicular
($\epsilon(k_1)\times\epsilon(k_2)$) polarization. The interest in the
ratio $R$ also crucially depends on the magnitude of the branching
ratio itself and is therefore important to examine the impact of QCD
corrections on both of them\footnote{One interesting implication of
this is that $A^+/A^-$ can be used to construct a CP-violating
observable which will pick up a dependence on
$Im(\lambda_t)/Re(\lambda_t)\sim O(\eta\lambda^2)$, where $\eta$ and
$\lambda$ correspond to the Wolfenstein parametrization of the CKM
matrix.}.

In the following we will use $f_{B_s}\!\simeq\! 200$ MeV,
$M_{B_s}\!=\! 5.37$ GeV, $m_b\!=\!4.8$ GeV, $m_c\!=\!1.5$ GeV,
$m_t\!=\!175$ GeV and $|\lambda_t|\!=\!|V_{tb}V_{ts}^*|\!=\!0.04$.
Using the experimental life time of the $B_s$ meson, $\tau_s\!=\!
1.61\cdot 10^{-12}$ s, we find that the branching ratio
$BR(B_s\rightarrow \gamma\gamma)$ goes, for $m_s\!=\!0.5$ GeV, from
$3.1\times 10^{-7}$ without QCD corrections to $5.0\times 10^{-7}$
with LO QCD corrections, therefore increasing by about $62\%$.  As far
as $A^+$ and $A^-$ are concerned, their ratio is substantially changed
by the acion of QCD corrections. It goes from $R\!=\!0.28$ without QCD
corrections to $R\!=\!0.55$ with LO QCD corrections. In fact at
$O(\alpha_s^0)$ both $A^+$ and $A^-$ depend on 1PR part of the
amplitude ($O_7$) and only $A^-$ is sensitive to the 1PI part ($O_2$).
When we switch on QCD corrections, the contribution of $O_7$ dominates
and drives $A^+$ and $A^-$ closer and closer. This effect is amplified
by the cancellation which takes place in the 1PI sector, mainly among
$O_2$ and $O_1$. 

The uncertainty on the perturbative calculation is dominated by the
scale-dependence of the LO Wilson coefficients, which is around
$25-30\%$. On the other hand, we estimate the uncertainty coming from
non-perturbative QCD effects, i.e. from the calculation of the matrix
element, to be of about $50\%$. Thus, attributing a $60\%$ uncertainty
to the central value ($5\times 10^{-7}$), we expect the branching
ratio to be about $2\!-\!8\times 10^{-7}$. It would be very useful to
have a more accurate calculation of these effects, perhaps by using
HQET beyond the leading order, so that a more precise theoretical
prediction can be obtained. Indeed it is not inconceivable that those
corrections will further increase the branching ratio for
$B_s\rightarrow\gamma\gamma$.

\section{Long distance QCD effects.}
\label{longdistance}

As far as the $B_s\rightarrow \gamma\gamma$ rare decay is concerned,
as we discussed in the previous Section, we expect long-distance QCD
corrections to be proportional to $1/m_s$ at the lowest order,
introducing an uncertainty that asks for a more accurate
computation of the matrix element. Other non perturbative effects
could come from the formation of $\bar c c$ bound states in the decay
process, i.e. from resonances. However in the $B_s\rightarrow
\gamma\gamma$ case these resonant states would be far off-shell and
they are not likely to give a significant contribution to the rate
(similar to the $b\rightarrow s\gamma$ case).

The inclusive decay $B\rightarrow X_s\gamma\gamma$ is, in this
respect, more problematic. In the region of invariant mass of the two
photons around $s\!\simeq\! 4 m_c^2/m_b^2$, the rate is going to be
dominated by the $\eta_c$ resonance, which subsequently decays into
two photons, i.e. by $b\rightarrow \eta_c s\rightarrow
s\gamma\gamma$. This could affect other regions of the spectrum and
constitute a serious problem. Moreover, we remind that in the
resonance region, the inclusive $B\rightarrow X_s\gamma\gamma$ decay
cannot be approximated anymore by the quark level process, as is the
case for $B\rightarrow X_s e^+e^-$ \cite{savage}. In order to
understand the relevance of our perturbative calculation we need to
include the resonance at the amplitude level and to estimate how it
affects the invariant mass ditribution, $d\Gamma/ds$, away from the
resonance peak. This will allow us to select those regions of the
spectrum which are free from major long-distance \emph{pollutions}.
In principle we should include in our analysis all the possible
resonant channels. However the $\eta_c$ resonance is dominant and is
enough to provide us with an idea of the resonant effects.

In order to model the contribution of the $\eta_c$ resonance we need
to provide an effective vertex both for the $b\rightarrow s\eta_c$
transition and for the $\eta_c\rightarrow\gamma\gamma$ decay that
follows it.  The $bs\eta_c$ vertex can be derived from the amplitude
for the $b\rightarrow s\eta_c$ decay \cite{desh}. Using the effective
hamiltonian in Eq.~(\ref{heff}) and parametrizing the axial vector
current matrix element

\be
\langle 0|\bar c\gamma^\mu\gamma_5 c|\eta_c\rangle = -i f_{\eta_c}
P_\eta^\mu
\ee

\noindent in terms of the decay constant $f_{\eta_c}\!\simeq\!300$ MeV
\cite{desh}, one gets\footnote{We assume simple factorization.}

\bea
\label{bsetacamplitude}
\langle s\eta_c|H_{\rm eff}|b\rangle &=&
-i\frac{G_F}{\sqrt{2}}\lambda_t f_{\eta_c}\left[C_1+C_3-C_5+
\frac{1}{N_c}(C_2+C_4-C_6)\right]\times\nonumber\\
&& \bar u_s \left[-m_s(1-\gamma_5)+
m_b(1+\gamma_5)\right]u_b\,\,\,.
\eea

\noindent For the values of the parameters used in this paper and
taking the LO Wilson coefficients from Table~\ref{tablecoeff}, we can
estimate $BR(b\rightarrow\eta_c+\mathrm{anything})\simeq 4\times
10^{-3}$, more restrictive than the present experimental upper bound
\cite{pdb}

\be
\label{betacanything}
BR(b\rightarrow\eta_c+\mathrm{anything})< 9\times 10^{-3}\,\,\,.
\ee
 
\noindent As far as the $\eta_c\gamma\gamma$ vertex is concerned, we
can assume the amplitude for $\eta_c\rightarrow\gamma\gamma$ to be of
the form

\be
A(\eta_c\rightarrow\gamma\gamma)= i B^- F_{\mu\nu} \tilde F^{\mu\nu}
\,\,\,,
\ee

\noindent and use the experimental measurement

\be
BR(\eta_c\rightarrow\gamma\gamma)=3\times 10^{-4}
\ee

\noindent to estimate $|B^-|\!\simeq\!2.5-3\times 10^{-3}$, for
$\Gamma_{\eta_c}\!=\!0.013$ GeV and $M_{\eta_c}\!\simeq\! 3$ GeV.  The
relative sign between the perturbative continuum and the resonant
contribution can be determined via the same kind of unitarity
arguments applied in Ref.~\cite{tung} to the $b\rightarrow s \psi$
case. In fact, in the resonance region the perturbative amplitude is
much smaller than the resonant one and therefore the relative sign
between the two terms of the amplitude has to be positive, as in
\cite{tung}.

The amplitude for the inclusive $b\rightarrow s\gamma\gamma$ decay can
now be written as the sum of a non-resonant, $A_{NR}$, and of a
resonant, $A_R$ part

\be
\label{amplres}
A(b\rightarrow s\gamma\gamma)=A_{NR}+A_R = A_{NR} +
 \left(-i\frac{e^2 G_F}{\sqrt{2}\pi^2}\lambda_t \right)
 \bar u_s(p^\prime)F_R W_R^{\mu\nu}u_b(p)\epsilon_\mu(k_1)
 \epsilon_\nu(k_2)
\,\,\,,
\ee

\noindent where $A_{NR}$, including LO QCD corrections, is given in
Eq.~(\ref{amplincl}) while $A_R$ has been expressed in terms of the
following coefficient and matrix element

\bea
F_R&=&i\frac{\pi}{4\alpha_e}f_{\eta_c}B^-
\left[C_1+C_3-C_5+\frac{1}{N_c}(C_2+C_4-C_6)\right]
\frac{1}{s-M_{\eta_c}^2+i\Gamma_{\eta_c}M_{\eta_c}}\,\,\,,\nonumber\\
W_R^{\mu\nu}&=&2i\epsilon^{\mu\nu\rho\sigma} k_{1\rho}k_{2\sigma}
\left[-m_s(1-\gamma_5)+m_b(1+\gamma_5)\right]\,\,\,.
\eea

\begin{figure}[ht]
\centering
\epsfxsize=4.in
\leavevmode\epsffile{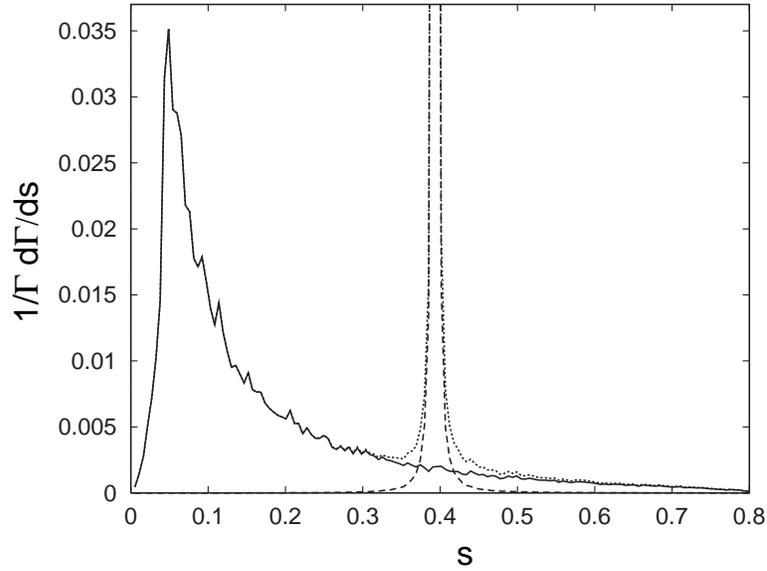}
\caption[]{The invariant mass distribution of the two photons in the
presence of the $\eta_c$ resonance, normalized to the total rate
$\Gamma_\mathrm{tot}\!=\!5.7\times 10^{-7}$, as obtained for
$m_s\!=\!0.5$ GeV. We show the pure non-resonant (solid), the pure
resonant (dot-dashed) and the total distribution (dashed). The
resonance peak is truncated in order to show the relevance of the
different contributions both inside and outside the resonance region.}
\label{s2res}
\end{figure}

\noindent If we use Eq.~(\ref{amplres}) to compute the invariant mass
distribution of the two photons, we see that the effect of the
resonance is very well localized around the resonance peak and does
not affect in particular the region for $s\!\le\!0.3$. We can define
in fact two regions, for $0.0\le s\le 0.3$ and for $s\ge 0.5$, in
which the effect of the $\eta_c$ resonance is practically negligible,
as one can see in Fig.~\ref{s2res}. 

\noindent Over these regions we can assume the validity of our 
perturbative calculation of Section \ref{inclusive} as well as of our
previous studies of the various kinematical distributions for
$b\rightarrow s\gamma\gamma$ decay \cite{bsgg}. Disregarding in the
perturbative calculation of Section \ref{inclusive} the contribution
of the resonance region, which we conservatively define as $0.3\le
s\le 0.5$, we find that the
\emph{perturbative} branching ratio is reduced by at most $14\%$. It
would be very useful to verify experimentally that the effect of the
$\eta_c$ resonance in the $B\rightarrow X_s\gamma\gamma$ case is not
so relevant, in comparison with what we know to be the case for
$B\rightarrow X_s e^+e^-$. In fact, if we consider the decay chain
$b\rightarrow s \psi$ followed by $\psi\rightarrow e^+e^-$ and use
both experimental \cite{pdb} and theoretical
\cite{burasreview,deshsanda} inputs, we can estimate that

\be
\frac{\Gamma(b\rightarrow s \psi)\Gamma(\psi\rightarrow e^+e^-)}
 {\Gamma(b\rightarrow s e^+e^-)}\simeq 1.4\times 10^2\,\,\,,
\ee

\noindent while the analogous quantity for $b\rightarrow s\eta_c$
followed by $\eta_c\rightarrow\gamma\gamma$ amounts to

\be
\frac{\Gamma(b\rightarrow s \eta_c)
\Gamma(\eta_c\rightarrow\gamma\gamma)}
 {\Gamma(b\rightarrow s\gamma\gamma)}\simeq 6\,\,\,.
\ee

\noindent This argument indirectly confirms the less dramatic impact
that the $\eta_c$ resonance has on the invariant mass distribution of
the two photons in the $b\rightarrow s\gamma\gamma$ decay.

\section*{Acknowledgments}

This research was supported in part by U.S. Department of Energy
contract DE-AC-76CH0016 (BNL).

\vspace{.5truecm} \emph{Note added.} While in the course of writing
this manuscript, we became aware of the following two papers: G.Hiller
and E.O. Iltan, hep-ph/9704385 and C.-H. V. Chang, G.-L. Lin and Y.-P.
Yao, hep-ph/9705345, in which the problem of QCD corrections as it
pertains only to the exclusive $B_s\rightarrow\gamma\gamma$ decay is
also discussed. We disagree with the first reference, where only the
contribution of $O_7$ and $O_2$ is considered. We agree with all the
results of the second reference, except for a few points that appear
to be misprints.

\appendix
\section{Study of the IR divergences of the rate.}
\label{ir}

In this appendix we want to show the explicit cancellation between the
IR singularities arising respectively in the $b\rightarrow
s\gamma\gamma$ rate from the bremstrahlung of a soft photon and in the
$O(\alpha_e)$ virtual corrections to the $b\rightarrow s\gamma$
amplitude. This will confirm our understanding of the endpoints of the
photon spectrum in the $b\rightarrow s\gamma\gamma$ decay. Our
calculation is very similar to what can be found in
Refs.~\cite{aligreub,pott} for the study of the gluon spectrum in
$b\rightarrow s\gamma g$. Many results could be taken from there
provided the different charge and color factors are adequately taken
into account. We have indeed reproduced the calculation we report in
this appendix and we can confirm\footnote{In the course of these
checks we came across a misprint in Eq.~(34) of Ref.~\cite{pott}. We are
very grateful to the author for confirming this. The correct
expression is given in Eq.~(\ref{dgamma7}).} \emph{a posteriori}
the results for $b\rightarrow s\gamma g$.

Moreover, as we already explained in Sec.~\ref{inclusive}, we are not
going to include $O(\alpha_e)$ virtual corrections to the
$b\rightarrow s\gamma$ amplitude in the calculation of the rate for
$b\rightarrow s\gamma\gamma$. In fact, we will just require the two
photons to be hard, imposing a minimun energy cut. Therefore, in the
present appendix we will consider only those aspects of the discussion
which are necessary to show the cancellation of the IR poles.

In the reaction $b(p)\rightarrow s(p^\prime)+\gamma(k_1)+\gamma(k_2)$,
the spectrum of any of the two photons presents two sharp singularities
in the vicinity of the endpoints, i.e. for $x_\gamma\rightarrow 0$ and
for $x_\gamma\rightarrow 1$, where we define
$x_\gamma\!=\!E_\gamma/E_\gamma^{\rm max}$ for $E_\gamma^{\rm
max}=(m_b^2-m_s^2)/2/m_b$. The variable $x_\gamma$
corresponds in general to the reduced energy of a given photon. 
To make contact with the notation introduced in (\ref{invariants}),
we can easily see that : 

\be
x_{\gamma_1}=\frac{u}{1-\rho}=\bar u\,\,\,\,\mbox{and}\,\,\,\,
x_{\gamma_2}=\frac{t}{1-\rho}=\bar t\,\,\,\,.
\ee

\noindent This singular behavior at the endpoints of the spectrum
corresponds to the presence of IR singularities in the rate for
$b\rightarrow s\gamma\gamma$, when the energy of one or the other of
two photons goes to zero, i.e. when $x_\gamma\rightarrow 0$ (the energy
of the photon under consideration) or $x_\gamma\rightarrow 1$ (the
energy of the other photon). 

These IR singularities originate from the integration of the
$A_{77}$ part of the square amplitude over the phase space of the two
photons.  As we can see from Eq.~(\ref{a77}), $A_{77}$ is symmetric
with respect to $(\bar u\leftrightarrow\bar t)$, i.e. under the
exchange of the two photons. Therefore the treatment of the two
endpoints is symmetric. Given the spectrum of one photon, we will
arbitrarily consider the endpoint $x_{\gamma_1}\rightarrow
1$. All our results will be valid in an analogous manner for the other
endpoint, i.e. for $x_{\gamma_2}\rightarrow 1$.

Let us consider the contribution of $O_7$ only to the differential
decay rate. Starting from Eq.~(\ref{dgamma}) and working out the
integration over the phase space in $D=4-2\epsilon$ dimension we get

\be
\label{dgamma7}
\frac{d\Gamma_7}{d\bar t\,d\bar u}=
 (1-\rho)^2 \frac{1}{4}\frac{\alpha_e}{\pi}\Gamma_0 F_7^2
 \frac{(1-\rho)^{-4\epsilon}(8\pi\mu^2/m_b^2)^{2\epsilon}}
 {\Gamma(2-2\epsilon)[\bar t\bar u (1-f(\bar t,\bar u)^2)^{1/2}]
 ^{2\epsilon}} A_{77}(\bar t,\bar u)\,\,\,,
\ee

\noindent where $A_{77}(\bar t,\bar u)$ is given in Eqs.~(\ref{aij})
and (\ref{a77}) and the function

\be
f(\bar t,\bar u)=1-\frac{2(\bar u+\bar t-1)}{(1-\rho)\bar t\bar u}
\ee

\noindent corresponds kinematically to the cosinus of the angle
between the two photons in the rest frame of the \emph{b} quark, when
expressed in terms of the invariants $\bar u$ and $\bar t$.
Moreover we denote by $\Gamma_0$ the quantity
\be
\label{gamma0}
\Gamma_0=\frac{G_F^2\alpha_e|\lambda_t|^2}{32\pi^4}m_b^5\,\,\,.
\ee

\noindent The origin of the singularity in $\bar u\rightarrow 1$
becomes evident after we integrate over $\bar t$ and similarly for the
singularity in $\bar t\rightarrow 1$ when we integrate over $\bar
u$\footnote{We could obtain this second singularity by looking, after
we integrate over $\bar t$, to the $\bar u\rightarrow 0$ endpoint of
the remaining integration. However, we prefer to use the symmetry
between the two photons.}. In particular, they are generated by the
the first bracket in $A_{77}^{(2)}$ (see Eq.~(\ref{a77})), whose
contribution, upon integration, reads

\be
\label{dgammair}
\frac{d\Gamma_{7,\bar u}^{\rm IR}}{d\bar u}=(1+\rho)(1-\rho)^3\frac{1}
{4} \frac{\alpha_e}{\pi} \Gamma_0 F_7^2 C_\epsilon 
 \left[-\frac{2 (G^{(a)}+\epsilon G^{(b)})}{(1-\bar u)^{1+2\epsilon}}
 \right]\,\,\,,
\ee

\noindent where 

\be 
C_\epsilon=\frac{(1-\rho)^{-4\epsilon}(4\pi\mu^2/m_b^2)^{2\epsilon}}
  {\Gamma(2-2\epsilon) \bar u^{2\epsilon}}\,\,\,,
\ee

\noindent and

\bea
\label{gfunctions}
G^{(a)} &=& \left(1+\frac{\rho}{1-u}\right)\bar u +
  \frac{1+\rho}{1-\rho}\log(1-u)\,\,\,,\\
G^{(b)} &=& \frac{\rho}{1-u}\left[2+\log(1-u)\right]\bar u-
 2\frac{1-u}{1-\rho}\log(1-u)+
 \frac{1+\rho}{1-\rho}\left[\frac{1}{2}\log^2(1-u)-2\mbox{Li}_2(u)
\right]
 \,\,\,,\nonumber
\eea

\noindent using the standard notation for the Spence function
$\mbox{Li}_2(x)=-\int_0^1 dt\log(1-xt)/t$. After the last integration
over $\bar u$, the IR singularity for $\bar u\rightarrow 1$ appears as
a pole in $\epsilon$, i.e.

\be
\label{gammairbremsu}
\Gamma_{7,\bar u\rightarrow 1}^\mathrm{IR}(b\rightarrow s\gamma\gamma)=
\int_0^1 d\bar u\frac{d\Gamma_{7,\bar u}^\mathrm{IR}}{d\bar u}=
(1+\rho)(1-\rho)^3\frac{1}{4}\frac{\alpha_e}{\pi}\Gamma_0 F_7^2
\frac{1}{\epsilon}\left[2+\frac{1+\rho}{1-\rho}\log\rho\right] \,\,+
\,\,\cdots\,\,\,,
\ee

\noindent where the dots indicate all other kinds of terms arising
from the integration. An analogous singularity arises for $\bar
t\rightarrow 1$ when we integrate the second term of $A_{77}^2$ (see
Eq.~(\ref{a77})) first over $d\bar u$ and then over $d\bar
t$. Therefore the rate has a total IR singularity given by

\be
\label{gammairbrems}
\Gamma_{7}^\mathrm{IR}(b\rightarrow s\gamma\gamma)=
\int_0^1 d\bar u\frac{d\Gamma_{7}^\mathrm{IR}}{d\bar u}=
(1+\rho)(1-\rho)^3\frac{1}{2}\frac{\alpha_e}{\pi}\Gamma_0 F_7^2
\frac{1}{\epsilon}\left[2+\frac{1+\rho}{1-\rho}\log\rho\right] \,\,+
\,\,O(1)\,\,\,,
\ee

\noindent where we have indicated with $O(1)$ all the other non
singular terms arising from the integration.

We will now show that the same IR singularity, but with opposite sign,
arises from the $O(\alpha_e)$ virtual corrections to the $b\rightarrow
s\gamma$ amplitude induced, of course, by the same operators $O_7$.
In this case, given the tree level $bs\gamma$ vertex induced by $O_7$,
we have to consider both self-energy and vertex $O(\alpha_e)$
corrections in the renormalized theory, i.e. taking into account the
wave function renormalization constants of the \emph{b} and of the
\emph{s} quark. The choice of gauge for the photon is not relevant if
the calculation is consistently performed (we checked the result in
both the Feynman and the Landau gauge) and the final result reads

\bea
\label{gamma7virt}
\Gamma_7^{(\alpha_e)}(b\rightarrow s\gamma)&=&
 (1+\rho)(1-\rho)^3\Gamma_0 F_7^2 (1+2 K_{\alpha_e})
 \frac{\Gamma(1-\epsilon)}{\Gamma(2-2\epsilon)}\left(\frac{4\pi}{m_b^2}
 \right)^\epsilon (1-\rho)^{-2\epsilon}\nonumber\\
&=&\,\,\,\,\ \Gamma_7^{(0)} + \delta\Gamma_7^{(\alpha_e)}\,\,\,,
\eea

\noindent where

\bea
\label{deltagamma7virt}
\Gamma_7^{(0)}(b\rightarrow s\gamma) &=& 
 (1+\rho)(1-\rho)^3\Gamma_0 F_7^2 \,\,\,,\\
\delta\Gamma_7^{(\alpha_e)}(b\rightarrow s\gamma) &=& 
 (1+\rho)(1-\rho)^3\Gamma_0 F_7^2 
 2 K_{\alpha_e}\frac{\Gamma(1-\epsilon)}{\Gamma(2-2\epsilon)}
 \left(\frac{4\pi}{m_b^2} \right)^\epsilon(1-\rho)^{-2\epsilon}
 \,\,\,,\nonumber
\eea

\noindent and

\be
\label{kappaalphae}
K_{\alpha_e} = \frac{1}{4}\frac{\alpha_e}{\pi} (4\pi)^\epsilon
 \Gamma(1+\epsilon)\left[
 -\frac{1}{\epsilon}\left(2+\frac{1+\rho}{1-\rho}\log\rho\right)
 \,\, + \,\, O(1)\right]\,\,\,.
\ee

\noindent It is now easy to verify that the pole terms cancel between 
Eq.~(\ref{gammairbrems}) and Eq.~(\ref{gamma7virt}), such that

\be
\Gamma_7^{\rm IR}(b\rightarrow s\gamma\gamma)+
 \delta\Gamma_7^{(\alpha_e)}(b\rightarrow s\gamma) = O(1)\,\,\,.
\ee

\noindent In the previous discussion we may have disregarded terms of 
$O(\epsilon)$ when they do not happen to multiply a quantity containing
$1/\epsilon$ poles and we have omitted all over a factor of 
$(m_b(m_b)/m_b)^2$ because it would not influence the cancellation of
the IR poles.

\end{document}